\ifx\LaTeXe\undefined
 old LaTeX version
  \documentstyle[12pt,epsfig]{article}
  \newcommand{\ifig}[1]{\mbox{\epsfig{file=#1,width=10cm,angle=270}}}
\else
\documentclass[12pt]{article}
\usepackage{a4}
\usepackage[dvips]{graphicx}
\usepackage{latexsym}
\usepackage{euscript}
\newcommand{\ifig}[1]{\includegraphics[height=9cm,width=14cm]{#1}}
\fi
\newcommand{\bc}{\begin{center}}
\newcommand{\ec}{\end{center}}
\newcommand{\be}{\begin{equation}}
\newcommand{\ee}{\end{equation}}
\newcommand{\bea}{\begin{eqnarray}}
\newcommand{\eea}{\end{eqnarray}}
\newcommand{\ba}{\begin{eqnarray}}
\newcommand{\ea}{\end{eqnarray}}
\newcommand{\amu}{A_{\mu}}
\newcommand{\m}{_{\mu}}

\newcommand{\simge}{\ \lower-
1.2pt\vbox{\hbox{\rlap{$>$}\lower5pt
\vbox{\hbox{$\sim$}}}}\ }

\newcommand{\AC} {{\cal{A}}}

\begin{document}
\pagestyle{empty} 
\vspace{-0.6in}
\begin{flushright}
BUHEP-99-27\\
ROME 1274/99 \\
\end{flushright}
\vskip 2.0in

\centerline{\large {\bf{Lattice Gauge Fixing for Parameter}}}
\centerline{\large {\bf{Dependent Covariant Gauges}}}
\vskip 1.0cm
\centerline{L. Giusti$^{(1)}$, M. L. Paciello$^{(2)}$, S. Petrarca$^{(2,3)}$,
B. Taglienti$^{(2)}$}
\centerline{\small $^1$ Boston University - Department of Physics, 590 Commonwealth Avenue} 
\centerline{Boston MA 02215 USA.}
\centerline{\small $^2$INFN, Sezione di Roma 1,
P.le A. Moro 2, I-00185 Roma, Italy.}
\centerline{\small  $^3$ Dipartimento di Fisica, Universit\`a di Roma "La
Sapienza",}
\centerline{\small P.le A. Moro 2, I-00185 Roma, Italy.}

\centerline{\small }
\vskip 1.0in
\begin{abstract}
We propose a non-perturbative procedure to fix 
generic covariant gauges on the lattice. Varying the gauge 
parameter, this gauge fixing provides a concrete method 
to check numerically the gauge dependence of correlators 
measured on the lattice. 
The new algorithm turns out to converge with a good efficiency.
As a preliminary physical result, we find a sensitive dependence of 
the gluon propagator on the gauge parameter.
\end{abstract}
\vfill
\pagestyle{empty}\clearpage
\setcounter{page}{1}
\pagestyle{plain}
\newpage 
\pagestyle{plain} \setcounter{page}{1}

\newpage

\section{Introduction}
Lattice gauge fixing is unavoidable to compute  
non perturbatively the propagators of the fundamental fields 
appearing in the QCD Lagrangian \cite{tutti_g}. It is 
also necessary in non-perturbative renormalization 
schemes~\cite{NPM,parrinello} which use gauge dependent matrix elements 
to renormalize composite operators, and it becomes a 
fundamental technical 
ingredient in the so called non gauge invariant quantizations of chiral 
theories~\cite{romaappr}.\\
Up to now, the Landau gauge is the only covariant gauge for which 
an efficient numerical algorithm is available.

A few years ago, an unconventional method to implement a generic covariant gauge 
on the lattice was proposed \cite{fachin}. 
In principle this procedure takes into account 
the contributions from each gauge orbit with appropriate weight.
With this method a gauge dependence of the gluon propagator, in particular
at zero momentum, had been found \cite{parrisoft}. On the other hand 
the implementation of this method for physical lattices can be numerically 
demanding. 

Here we follow a more conservative procedure to fix a generic covariant gauge 
on the lattice that stems from ref.~\cite{giusti}. It is based, as in the 
Landau case, on the minimization of a functional $H_A[G]$
chosen in such a way that its absolute
minima correspond to a gauge transformation $G$ 
satisfying the appropriate gauge condition (see section \ref{sec:LGF}).
In the continuum it is easy to show \cite{books} that 
this procedure is equivalent to the Faddeev-Popov quantization for 
covariant gauges, i.e. the two procedures lead to the same matrix elements 
for a  generic gauge dependent operator. As usual the 
Faddeev-Popov factor can be 
written as a Gaussian integral of local Grassman variables, the 
resulting effective action is invariant under the 
BRST transformations and the correlation 
functions of the operators satisfy the appropriate 
Slavnov-Taylor identities.
On the compact lattice the presence of Gribov copies is an obstruction
to define  partition functions with exact BRST invariance 
as it has been shown by H. Neuberger
a few years ago \cite{herbert}. The Neuberger's argument 
concerns a fundamental theoretical point whose relevance for 
the measurement of gauge dependent operators  on the lattice 
must be clarified.
Nevertheless, we believe that  new numerical gauge fixing algorithms
 represent a concrete tool to improve the understanding of the lattice 
 gauge properties.
   
In the following
we propose a new numerical gauge fixing procedure 
which generalizes the usual
Landau gauge fixing to generic covariant gauge.
The numerical implementation of our procedure on the lattice 
uses a straight-forward generalization of the 
standard Landau algorithm, i.e. a steepest descent iterative algorithm 
minimizes a discretization  of the new functional $H_A[G]$. 
In order to have an efficient and feasible gauge fixing procedure, 
the discretized functional $H_U[G]$ 
must be chosen carefully. In fact a na{\"\i}ve discretization of this 
functional, due to its complicated 
structure, would lead to an algorithm which 
either does not converge or takes too much computer time 
\cite{pisa}.

The main purpose of this paper is to show that a 
simple discretization $H_U[G]$ exists and leads to a 
minimization algorithm which converges with a good efficiency. 
The application of this algorithm 
to the non-perturbative evaluation of gauge dependent Green
functions like the quark and gluon propagator 
will clarify the gauge dependence of the fitted parameters. 

In order to show the feasibility of our method,
we have computed the gluon propagator at different
values of the gauge parameter at small volumes, finding a sensitive 
gauge dependence. 

The plan of the paper is as follows. In section \ref{sec:LGF} we review the 
method proposed to fix a generic covariant gauge on the lattice. 
In section \ref{sec:driven} we describe a very simple and efficient 
discretization of the gauge fixing functional. In section \ref{sec:results} 
we give the details of our numerical simulations and report our main 
numerical results. 

\section{Covariant Gauge Fixing}\label{sec:LGF}
In this section we set the notation and briefly formulate the 
covariant gauge fixing method in the continuum.
In the following we will neglect the problem of Gribov copies,
assuming that a gauge section in the space of gauge fields
intersects all gauge orbits once and only once. 
In the Landau gauge the expectation value of a gauge dependent 
operator 
is given by
\be
\langle {\cal O} \rangle = 
\int \delta\amu \delta \eta \delta \bar{\eta}\, {\cal O}\, e^{-S(A)-
S_{ghost}(\eta,\bar{\eta},A)} \delta(\partial\m\amu)
\; ,\nonumber
\ee
and the gauge fixing condition can be enforced non-perturbatively by 
minimizing Gribov's functional \cite{Gribov} 
\be\label{eq:landau}
F_A[G]\equiv ||A^G||^2=
\int\mbox{\rm Tr}\left(A^G_{\mu}A^G_{\mu}\right)d^4x \; .
\ee
The Landau gauge is
readily extended to a general covariant gauge-fixing condition of the form
\be\label{dinamic1}
\partial\m\amu^G(x)=\Lambda(x)\; , 
\ee
where $\Lambda(x)=\lambda^a(x) \frac{T^a}{2}$ belongs to the Lie algebra of the
group and $\mbox{\rm Tr}(T^aT^b)=2\delta^{ab}$.
Since gauge-invariant quantities are not sensitive to changes of
gauge condition, it is possible to average over $\Lambda(x)$ with a
Gaussian weight 
\be\label{o2}
\langle {\cal O} \rangle =\int\delta\Lambda 
e^{-\frac{1}{\alpha}\int d^4x Tr[\Lambda^2]}
\int \delta\amu \delta\eta \delta\bar{\eta}\, {\cal O}\, e^{-S(A)-
S_{ghost}(\eta,\bar{\eta},A)}
\delta(\partial\m\amu-\Lambda)\; ,
\ee
obtaining the standard formula
\be\label{o3}
\langle {\cal O}\rangle =\int \delta\amu \delta \eta \delta \bar{\eta}\, 
{\cal O}\, e^{-S(A)-
S_{ghost}(\eta,\bar{\eta},A)}
e^{-\frac{1}{\alpha}\int d^4x Tr [(\partial_{\mu}A_{\mu})^2]}\; .\nonumber
\ee
This formula is adopted as a definition for the expectation value
of gauge dependent operators. The invariance under 
BRST transformations of the effective action in eq.~(\ref{o3}) leads 
to the Slavnov-Taylor identities among different correlation functions. 
In particular the longitudinal part of the gluon propagator 
has to be equal to the free one, i.e. \cite{books}
\be\label{eq:STIC}
\frac{2}{N^2-1}
\mbox{\rm Tr} \langle \partial_\mu A_\mu(x) 
\partial_\nu A_\nu(y) \rangle \propto \alpha\delta(x-y) \; ,
\ee
where $N$ is the number of colors. 
Following the usual technique, the gauge-fixing condition (\ref{dinamic1})
is obtained non perturbatively by minimizing the new functional~\cite{giusti} 
\be\label{cov11}
H_A[G]\equiv 
\int d^4x\mbox{\rm Tr}\left[(\partial_{\mu}A^G_{\mu}-\Lambda)  
(\partial_{\nu}A^G_{\nu}-\Lambda)\right]\; , 
\ee
which obviously reaches its absolute minima ($H_A[G]=0$) 
when eq.~(\ref{dinamic1}) is satisfied.
Therefore in this case the Gribov copies
of the equation (\ref{dinamic1}) are associated with
different absolute minima of eq.~(\ref{cov11}). 
All the stationary points of $H_A[G]$ correspond to the following 
gauge condition 
\be\label{eq:brutta}
D_\nu \partial_\nu(\partial_\mu A^G_\mu-\Lambda) = 0\; ,
\ee 
where $D_\nu$ is the covariant derivative.
Hence, due to the complexity of the new functional, its minimization could
find  ``spurious'' solutions which  correspond to zero modes of the operator
$D_\nu \partial_\nu$ and do not 
satisfy the gauge condition in eq.~(\ref{dinamic1}).
Of course the numerical minimization of the discretized version of 
eq.~(\ref{cov11}) can reach relative minima (spurious solutions) with
$H_A[G]\rightarrow 0$ which are not distinguishable from the absolute
minima. Hence  this could simulate the effect of 
an enlarged set of numerical Gribov copies. Within the 
limitations of this preliminary study we do not find any 
practical difference
between the  use of the new functional with respect to the standard
Landau one (see results given in Fig.~\ref{fig:aiai} below).

On the lattice, the expectation value of a gauge dependent operator 
${\cal O}$ in a generic covariant gauge is
\be\label{omedio}
\langle{\cal O}\rangle=\frac{1}{Z}\int d\Lambda
e^{-\frac{1}{\alpha}\sum_{x} Tr [\Lambda^2]} 
\int dU {\cal O}({U^{G_\alpha}})
e^{-\beta S(U)} \; ,
\ee
which is the straight-forward discretization of Eq.~(\ref{o2}),
where $\Lambda$ is dimensionless on the lattice.
$S(U)$ is the Wilson lattice gauge invariant action 
and $G_\alpha$ is the gauge transformation that 
minimizes the discretized version of the functional (\ref{cov11}).
On the lattice, the correct
adjustment to the measure is included in eq.~(\ref{omedio}) 
by evaluating the operator over the gauge rotated links.
Therefore it is not necessary to introduce ghost fields 
but it is mandatory to fulfill the gauge fixing condition numerically in 
order to get $G_\alpha$.

\section{The Driven Discretization}\label{sec:driven}
In the Wilson discretization of gauge theories, the fundamental fields
are the links $U_\mu$ which act as parallel transporters of the theory.
Hence the lattice fields $A_\mu$  are derived quantities which 
tend to the continuum gluon field as the lattice spacing 
vanishes. As a consequence on the lattice it is possible to choose different
definitions of $A_\mu$ formally equal up $O(a)$ terms. 
In quantum field theory this ambiguity 
is well understood because 
any pair of operators, differing from
each other by irrelevant terms, will tend to the same continuum 
operator. This feature, checked in perturbation theory,
has been verified numerically 
at the non-perturbative level in ref.~\cite{noigauge},
where it has been shown that different definitions
of the gluon field give rise to Green's functions
proportional to each other, guaranteeing the uniqueness of the 
continuum gluon field. 

The freedom to choose the lattice definition of 
$A_\mu$ can be used to build discretized functionals
which lead to efficient gauge-fixing algorithms. 
In the standard Landau gauge fixing, for example,  
the discretization of the functional
(\ref{eq:landau}) is given by 
\be
F_U[G] = - \frac{1}{VT a^2 g^2} Tr \sum_{x,\mu} \left[ U^G_\mu(x) + 
U^{G\dagger}_\mu(x) - 2 I \right]\; , 
\ee
where $V$ is the 3-dimensional volume and $T$ the time size of the lattice.
This formula corresponds, only up to $O(a)$ terms, 
to the na{\"\i}ve discretization 
of eq.~(\ref{eq:landau}) that is obtained from the standard lattice 
definition of the gluon field
\be
A_{\mu} (x) 
\equiv  \left[{{U_{\mu} (x) - U_{\mu}^{\dagger} (x)}\over
{2 i a g}}\right]_{Traceless}\; . 
\label{eq:prima}
\ee  
On the other hand $F_U[G]$ has the important property 
that it depends only 
linearly on 
$G(\bar x)$, when the iterative algorithm 
visits the lattice point $\bar x$. This feature 
would be spoiled if $F_U[G]$ were defined assuming literally the na{\"\i}ve 
discretization (\ref{eq:prima}) in eq.~(\ref{eq:landau}). 

In order to study the convergence of the
algorithm, two quantities are usually monitored as a function of the number of
iteration steps: $F_U[G]$ itself and
\be\label{eq:theta_Fold}
\theta_F = \frac{1}{VT} \sum_x Tr [\Delta_F\Delta_F^{\dagger}]\; , 
\ee 
where
\ba\label{eq:deltaF}
\Delta_F(x)  = 
\left[ X_F(x) - X_F^{\dagger}(x) \right]_{Traceless}\; 
\propto \frac{\delta F_U[G]}{\delta \epsilon}, 
\ea 
being $G=e^{i\epsilon^a \frac{T^a}{2}}$ and 
\ba\label{eq:X_1}
X_F(x) & = &  \sum_\mu \left(U_\mu(x) + U^\dagger_\mu(x-\mu)\right)\; .
\ea
$\Delta_F(x)$ is proportional to the first derivative of $F_U[G]$ 
and reaches zero as the functional is extremized. Therefore
the quality of the gauge fixing is determined by the  
parameter $\theta_F$ which corresponds to the continuum quantity 
$\int d^4x\mbox{\rm Tr} (\partial_{\mu}A_{\mu})^2 $.

A na{\"\i}ve discretization of $H_A[G]$ will generate a quadratic
dependence on $G$, which could prevent the convergence of the algorithm.
This obstacle has been overcome by taking advantage of the freedom to 
choose the gluon field definition. In fact, as in the Landau case,
it has been possible to find a discretization of $H_A[G]$ 
("driven discretization") that depends linearly on $G ( {\bar x})$ 
and corresponds, up to $O(a)$ terms, to the continuum limit, eq. (\ref{cov11}).
This aim can be
reached by choosing each different term of $H_U[G]$ in order to
guarantee the local linear dependence on $G({\bar  x})$, instead of
being the algebraic consequence of a particular $A_\mu$ definition. 
We propose the following compact form of $H_U[G]$: 
\be
H_U[G]  =  \frac{1}{VT a^4 g^2} Tr \sum_{x} J^G(x) 
J^{G \dagger }(x) \; , \label{eq:HJJ}
\ee
where 
\ba
J(x)  & = &  N(x) -i g \Lambda(x)\; , \nonumber\\
N(x)  & = &  -8 I + \sum_{\nu} \left( U_\nu^\dagger(x-\nu) + U_\nu(x)
\right)\; .
\ea
It is easy to see that locally $H_U[G]$ transforms linearly in $G({\bar
x})$ and its continuum limit is the functional 
(\ref{cov11}). $H_U[G]$ is positive semidefinite and,
unlike the Landau case, it is not invariant 
under global gauge transformations.\\
The functional $H_U[G]$ can be 
minimized using the same numerical technique
adopted in the Landau case. In order to study the convergence of the
algorithm, two quantities can be monitored as a function of the number of
iteration steps: $H_U[G]$ itself and
\be\label{eq:theta_Hnew2}
\theta_H = \frac{1}{VT} \sum_x Tr [\Delta_H\Delta_H^\dagger]\; ,
\ee 
where
\ba\label{eq:deltaHnew2}
\Delta_H(x) &  = & \left[ X_H(x) - X_H^{\dagger}(x) \right]_{Traceless}
\propto \frac{\delta H_U[G]}{\delta \epsilon}\; 
\ea 
and 
\ba\label{eq:X_2}
X_H(x) & = &  \sum_\mu \left(U_\mu(x)J(x+\mu) + 
U^\dagger_\mu(x-\mu)J(x-\mu) \right)\nonumber\\ 
& - & 8 J(x) - 72 I 
+ig N(x) \Lambda(x) \; . 
\ea
$\Delta_H$ is the driven discretization of the eq.~(\ref{eq:brutta}); it is 
proportional to the first derivative of $H_U[G]$ and,
analogously to the continuum, it is invariant under the transformations 
$\Lambda(x) \rightarrow \Lambda(x) + C$, where $C$ is a constant matrix
belonging to the $SU(3)$  algebra. During the minimization process 
$\theta_H$ decreases to zero and $H_U[G]$ becomes constant. 
The quality of the convergence is measured by the final value of $\theta_H$.

\section{Numerical Simulations and Results}\label{sec:results}
We have generated 50 $SU(3)$ thermalized link configurations using the Wilson 
action with periodic boundary conditions at $\beta=6.0$ for $8^4$ and
$8^3\times 16$ volumes. Following the prescription contained in 
eq.~(\ref{omedio}), for each Monte Carlo configuration and for each lattice 
site we have extracted a matrix $\Lambda(x)$ 
according to a Gaussian distribution at a fixed $\alpha$ value. 
The gauge-fixing code implements an 
iterative overrelaxed minimization algorithm for $F_U[G]$ and 
$H_U[G]$. We have monitored the quantities $F_U$ and $\theta_F$ for the 
standard Landau gauge-fixing algorithm  and 
$H_U[G]$ and $\theta_H$ for the new one after every lattice sweep. 
We have found that the value
of the overrelaxing parameter $\omega$ adopted for the Landau algorithm
$\omega=1.72$ is  a good choice also for the new gauge fixing.
 
\begin{figure}[h]
\bc
\ifig{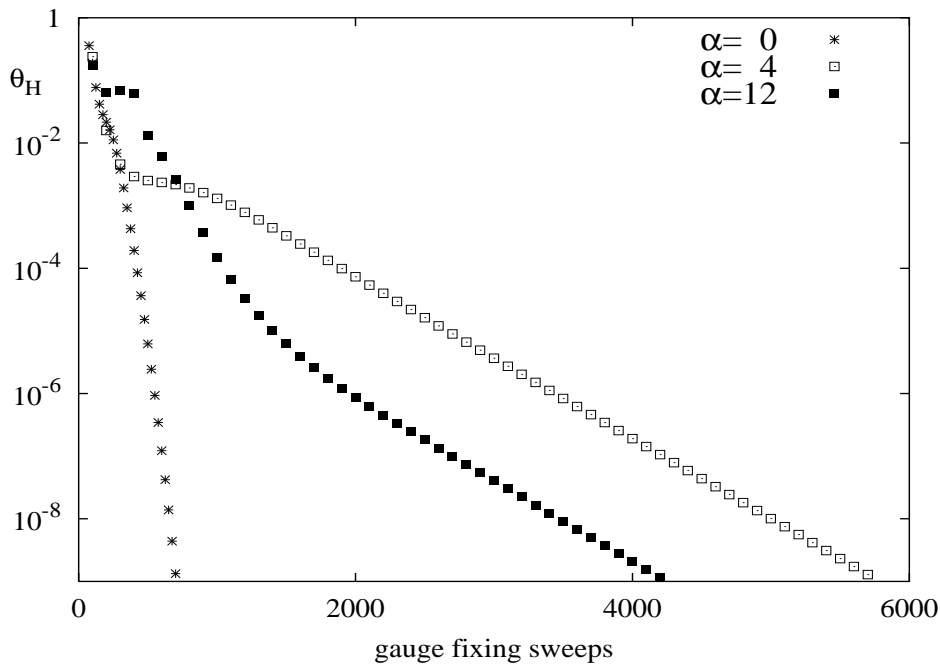}
\caption{\small{Typical behaviour of $\theta_H$ vs gauge fixing sweeps 
for different
choices of $\alpha$, for $\beta=6.0$ and $V\cdot T=8^4$.}}
\label{fig:theta}
\ec
\end{figure}

In Fig.~\ref{fig:theta} we report the values of $\theta_H$
for different $\alpha$ values as a function of the gauge fixing
sweeps for a typical thermalized configuration at $\beta=6.0$ and 
$V\cdot T=8^4$.
Each sweep of the new algorithm takes $\sim 15\%$
more computer time with respect to the Landau case. 
For $\alpha=0$, which in 
this new procedure corresponds to the Landau gauge and is obtained 
by taking $\Lambda=0$, the number of sweeps 
necessary to fix the gauge is $\simeq 1.2$  
times that of the standard algorithm. 
In the case of $\alpha\neq 0$, the number of sweeps
to minimize $H_U[G]$ with a given precision
increases when $\alpha$ decreases.
 
Once the configuration have been rotated in a given gauge, we 
have tested our procedure computing
the following two point correlation functions
\ba
\langle {\cal A}_0\,{\cal A}_0\rangle (t) &\equiv & \frac{1}{V^2}
\sum_{{\bf x,y}} \mbox{Tr} \langle  A_0({\bf x},t)A_0({\bf y},0) \rangle \; , 
\label{eq:A0A0}\\
\langle {\cal A}_i\,{\cal A}_i\rangle (t) &\equiv & \frac{1}{3 V^2}  
 \sum_{i}\sum_{{\bf x,y}} \mbox{Tr} \langle  A_i({\bf x},t)A_i({\bf y},0)\rangle
 \; , \label{eq:AiAi}\\
\langle \partial {\cal A}\, \partial {\cal A}\rangle (t) &\equiv & \frac{1}{4 V}  
 \sum_{\mu,\nu}\sum_{{\bf x}} \mbox{Tr} \langle  \partial_\mu A_\mu({\bf x},t) 
 \partial_\nu A_\nu({\bf x},0)\rangle \; ,\label{eq:dAdA}
\ea
where $\mu$ and $\nu$ run from 1 to 4, $i$ from 1 to 3 and the 
trace is over the color indices. 
In eqs.~(\ref{eq:A0A0})-(\ref{eq:dAdA}) the definition (\ref{eq:prima})
is adopted for the 
gluon field on the lattice and $\partial_\mu$ indicates 
the usual backward derivative.
The correlators in eqs.~(\ref{eq:A0A0}) and (\ref{eq:AiAi}) are relevant
to the investigation of the QCD gluon sector, while we use the correlation 
in eq.~(\ref{eq:dAdA}) to check the Slavnov-Taylor identity for the 
longitudinal component of the gluon propagator. The statistical errors 
for the correlation functions have been estimated by the jacknife 
method.
A thorough study of these operators will be 
presented in a forthcoming paper \cite{teorico}.

\begin{figure}[h]
\bc
\ifig{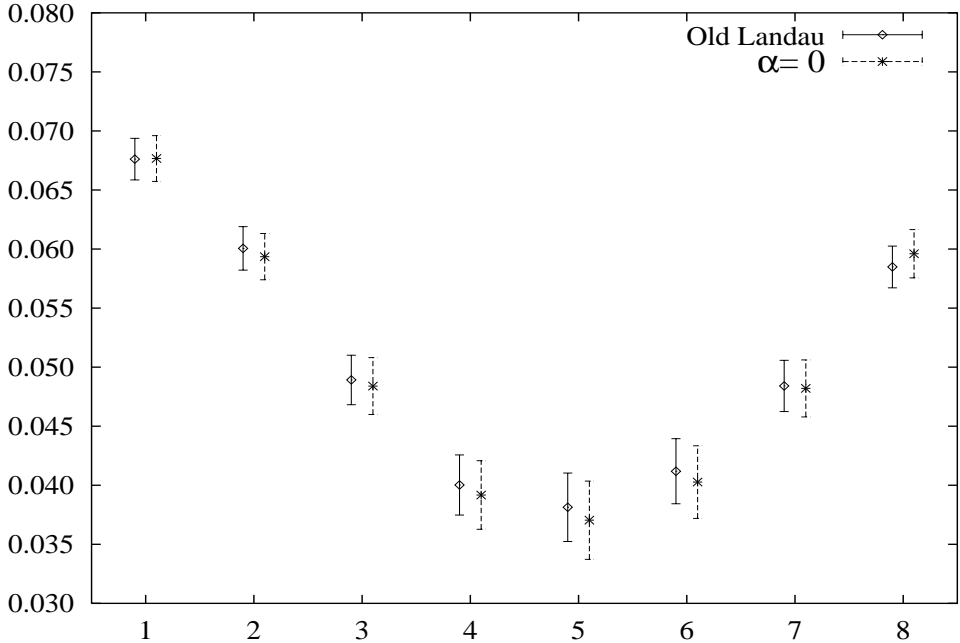}
\caption[]{\small{Comparison between   
$\langle {\AC}_i {\AC}_i\rangle_F$ (see eq.~(\ref{eq:AiAi})) obtained with 
the old 
functional $F_U[G]$ 
(circles) and $\langle {\AC}_i {\AC}_i\rangle_H$ obtained 
with the new one $H_U[G]$
with $\Lambda=0$ (crosses)  as  
function of time $t$ for a set of $50$ thermalized configurations 
at $\beta=6.0$ with a  
volume $V\cdot T=8^4$. The errors are jacknife and the data have been
slightly displaced in $t$ for clarity.}}
\label{fig:aiai}
\ec
\end{figure}

In Fig.~\ref{fig:aiai} the Green
functions computed with the standard Landau gauge fixing
$\langle {\cal A}_i {\cal A}_i \rangle_F$ and 
$\langle {\cal A}_i {\cal A}_i \rangle_H$ evaluated with the new algorithm 
with $\alpha=0$, are reported. 
The remarkable agreement shows that the new algorithm reproduces
very well the 
results of the standard one  within an overall scale factor close 
to 1.
Moreover, the possible spurious solutions of the 
new gauge condition apparently do not 
affect this operator at least at $\alpha=0$.

Although in this paper we did not perform a systematic study of the Gribov
copies for this gauge fixing procedure,
a rough preliminary search shows, for $\alpha=0$, the same pattern of copies 
of the Landau case.

\begin{figure}[h]
\bc
\ifig{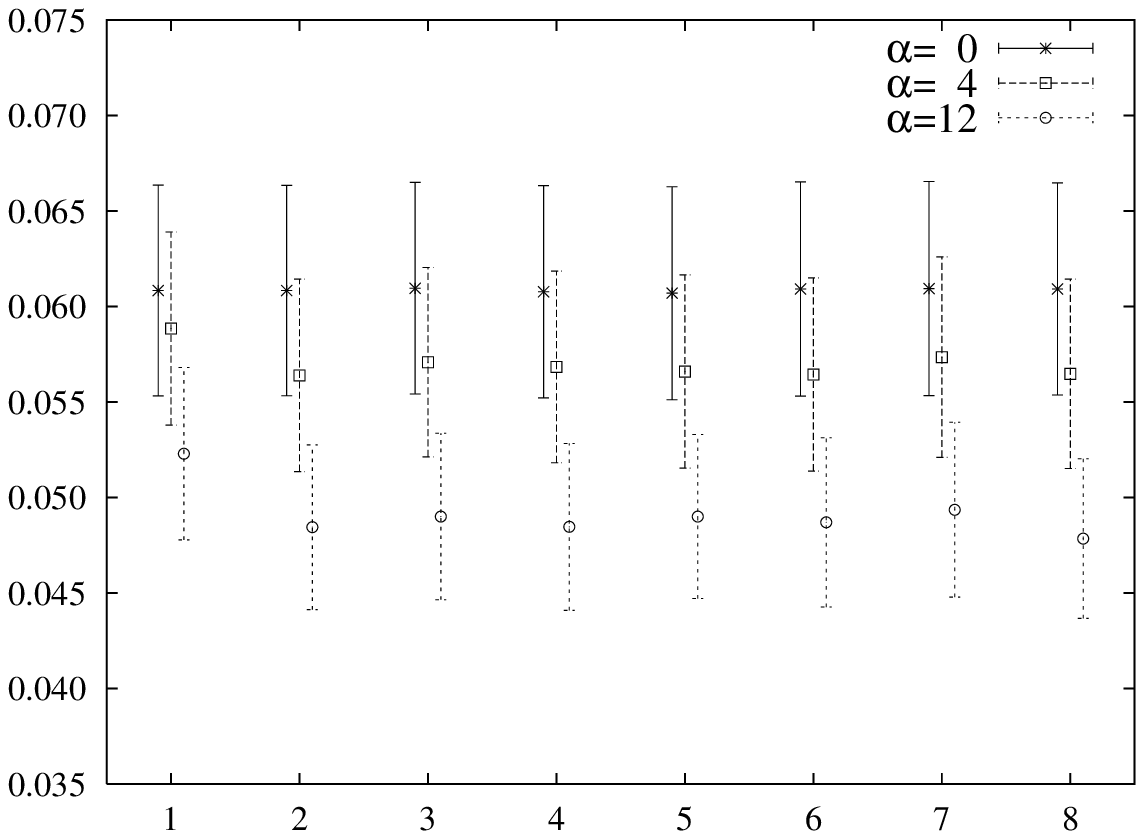}
\caption{\small{Behaviour of $\langle {\cal A}_0 {\cal A}_0\rangle_H$
(see eq.~(\ref{eq:A0A0})) 
obtained with the new functional $H_U[G]$ for different $\alpha$
values as  
function of time $t$ for a set of $50$ thermalized configurations 
at $\beta=6.0$ with a  
volume $V\cdot T=8^4$. The errors are jacknife and the data have been
slightly displaced in $t$ for clarity.}}
\label{fig:a0a0}
\ec
\end{figure}

In Fig.~\ref{fig:a0a0} the data for the correlation 
(\ref{eq:A0A0}) are reported. 
For $\alpha=0$ they show the well known flat behaviour 
forced by the Landau gauge.
For $\alpha\neq 0$ the data show an almost flat curve with
an enhancement at $t=1$ probably due to the effect of contact terms.
In this case the flatness stems from the fact that 
$\int d^3 x \partial_0 A_0 \simeq 0$ because 
the average value of $\Lambda$ on each time slice  
is negligible.

\begin{figure}[h]
\bc
\ifig{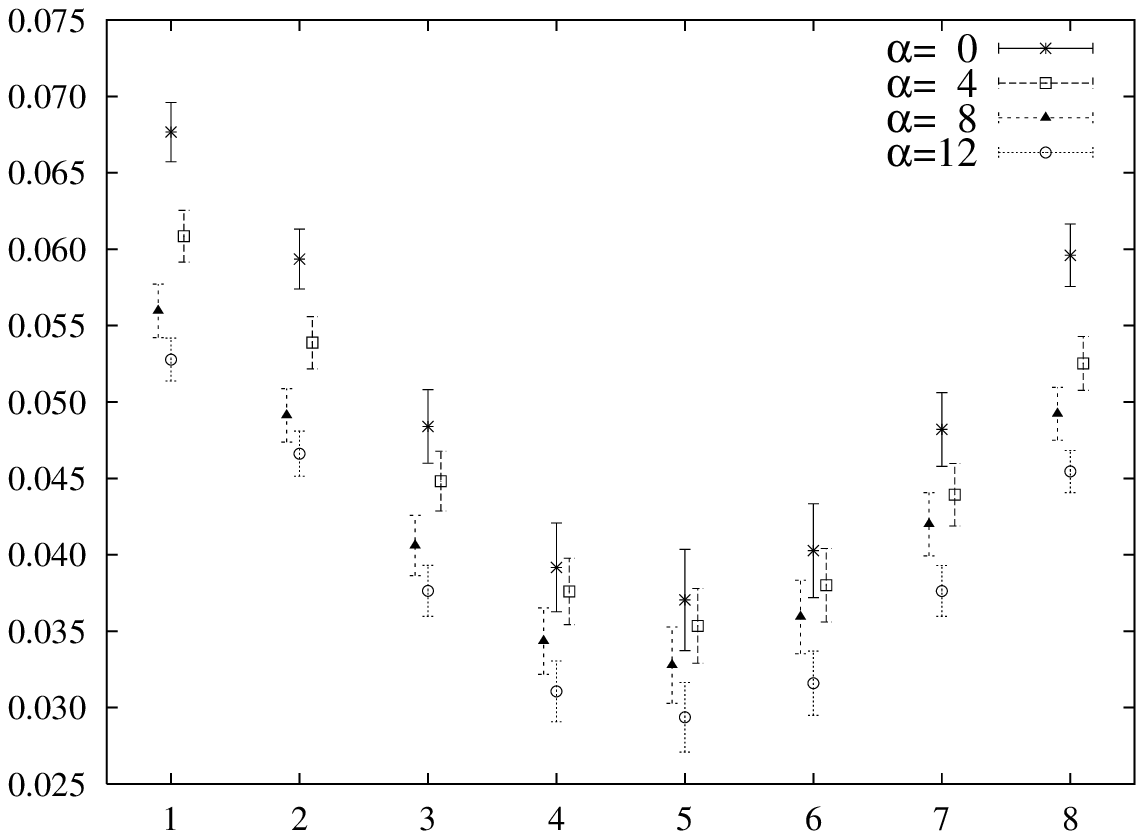}
\caption{\small{Behaviour of $\langle {\cal A}_i {\cal A}_i\rangle_H$
(see eq.~(\ref{eq:AiAi})) 
obtained with the new functional $H_U[G]$ for different $\alpha$
values as  
function of time $t$ for a set of $50$ thermalized configurations 
at $\beta=6.0$ with a  
volume $V\cdot T=8^4$. 
The errors are jacknife and the data have been
slightly displaced in $t$ for clarity.}}
\label{fig:aiai-alfa}
\ec
\end{figure}

In Fig.~\ref{fig:aiai-alfa} our results for 
the correlations $\langle {\cal A}_i {\cal A}_i\rangle_H$ for different 
$\alpha$ values as a function of the time slice $t$ are reported for 
a set of $50$ SU(3) configurations at $\beta=6.0$ and $V\cdot T=8^4$.

Although in the case of small volume and a small number 
of configurations, the gauge dependence of the gluon propagator 
is clearly shown. For increasing $\alpha$ values, 
the time dependence of the gluon propagator becomes flatter   
since it approaches a limit where the gauge fixing effects disappear.
It is also interesting to note that the $\alpha$ dependence
of the gluon propagator shown in Fig.~{\ref{fig:aiai-alfa}} does not 
seem to be re-absorbed by an overall scaling factor.
This plot shows the feasibility of this procedure to 
study the gauge dependence of physically interesting 
correlators.

We complete the numerical
checks of the gauge-fixing we are proposing, showing the
numerical behaviour  of a particular
Slavnov-Taylor identity. This identity is a direct consequence
of our way to implement the covariant gauge fixing  and
it must be understood as a numerical check of the  procedure.
This check  measures the correlator of the operator 
${\partial {\cal A}}$ with itself,
 and if this operator has been well gauge-fixed, it 
is related to the autocorrelator of the $\Lambda$'s and then it must be
a delta function.
In fact,
for a finite volume $V\cdot T$ with periodic boundary conditions, 
the na{\"\i}ve Slavnov-Taylor identity gives for the correlator of eq.~(\ref{eq:dAdA})
the following relation:
\be\label{eq:STI2}
\langle \partial {\cal A}\, \partial {\cal A}\rangle (t)
\propto \alpha (\delta_{t,0} - \frac{1}{V \cdot T}) + {\cal O}(a)\; .
\ee
\begin{figure}[h]
\bc
\ifig{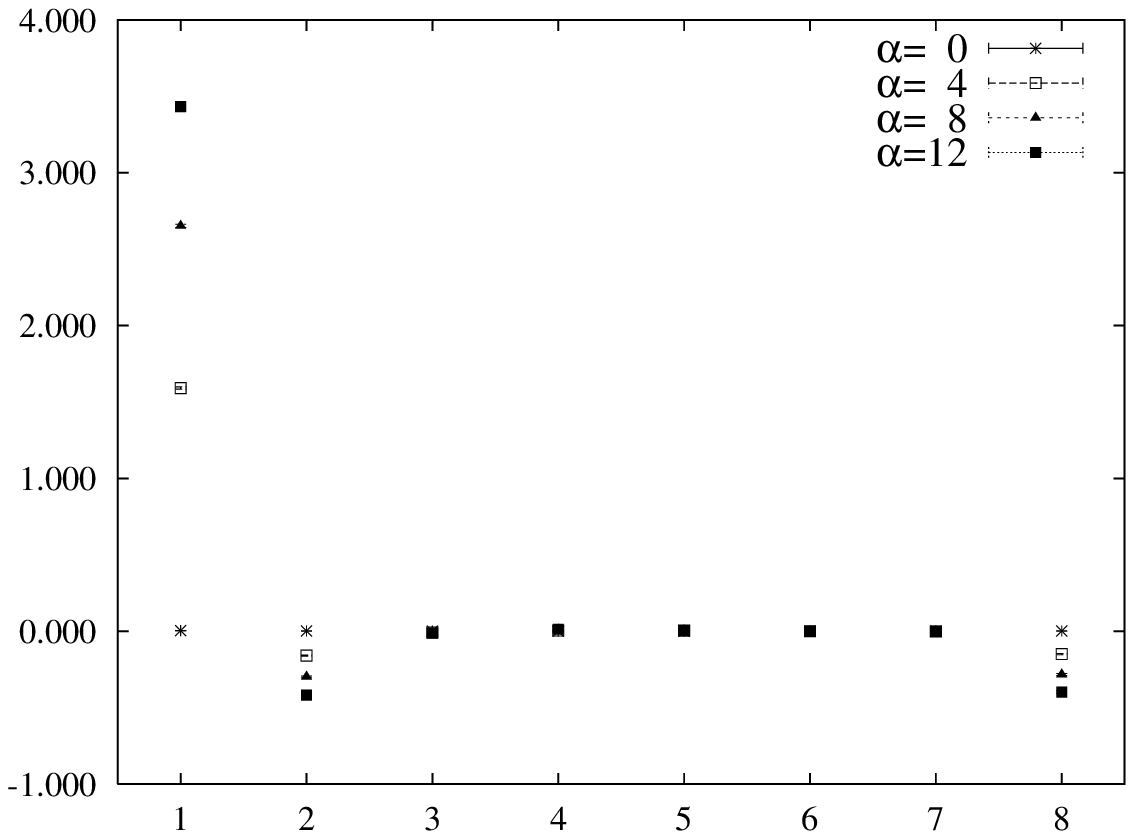}
\caption{\small{Behaviour of 
$\langle \partial {\cal A} \partial {\cal A}\rangle_H$ 
(see eq.~(\ref{eq:dAdA})) 
obtained with the new functional $H_U[G]$ for different $\alpha$
values as  
function of time $t$ for a set of $50$ thermalized configurations 
at $\beta=6.0$ with a volume $V\cdot T=8^4$. The errors are jacknife.}}
\label{fig:dito}
\ec
\end{figure}

Eq.~(\ref{eq:STI2}) 
shows that the 
shape of the correlation function (\ref{eq:dAdA}) is 
expected to be proportional to a 
$\delta$-function up to the square of the $\partial_\mu A_\mu(x)$ renormalization
constant, apart from volume and ${\cal O}(a)$ effects.
In order to verify eq. (\ref{eq:STI2}), in Fig.~\ref{fig:dito} 
we plot the correlation functions (\ref{eq:dAdA}) versus
the time $t$. 
We believe that
the good agreement between these data and eq.~(\ref{eq:STI2}) is an 
indication of the absence of possible spurious
solutions
induced by our method.
The small signal at $t=2,8$ is likely due to 
the residual correlation induced by the enhancement in the 
point $t=1$. 

\section{Conclusions}
In this paper we have described an efficient method to 
fix non-perturbatively a generic covariant gauge 
on the lattice. This procedure is equivalent to the Faddeev-Popov
quantization for covariant gauges. Therefore it 
can be used to compare the lattice renormalized correlation functions
with the same quantities  computed in continuum perturbation theory.\\
We have tested the algorithm on different SU(3) lattices for different 
values of the gauge parameter $\alpha$. Our numerical data show 
that the statistical fluctuations of measured quantities 
is comparable to the standard Landau gauge fixing algorithm.\\
We have computed 
the correlation functions relevant for the investigations of the gluon 
propagator at a few $\alpha$ values and we have found a 
sensitive dependence on the gauge parameter.\\
In collaboration with the Boston University Center for Computational 
Science, we are applying this method to the study of the gluon propagator on
physical lattices. 
\section*{Acknowledgments}
We warmly thank Massimo Testa for many fruitful discussions 
on this subject.

\noindent L.G. thanks Claudio Rebbi for interesting discussions on this subject.

\noindent We thank the Center for Computational Science of Boston 
University where part of this computation has been done.

\noindent L. G. was  supported in part under DOE grant DE-FG02-91ER40676.
 
\noindent S.~P. has been partially supported by the INFN-MIT ``Bruno Rossi'' 
Exchange Program while part of this paper was
written. He thanks all the members of the MIT Center for 
Theoretical Physics for their warm hospitality. 
   

\begin{thebibliography}{99}
\bibitem{tutti_g}
For the gluon propagator see the recent review by 
J.~E.~Mandula, hep-lat/9907020 and reference therein.
For the quark propagator see D.~Becirevic, V.~Gimenez, 
V.~Lubicz and G.~Martinelli,
hep-lat/9909082 and reference therein.
\bibitem{NPM}
G. Martinelli, C. Pittori, C. T. Sachrajda, M. Testa, A.~Vladikas,
Nucl. Phys. {\bf B445} (1995) 81.
\bibitem{parrinello}
B. Alles, D. S. Henty, H. Panagopoulos, C. Parrinello, 
C. Pittori and  D. G. Richards,
Nucl. Phys. {\bf B502} (1997) 325. 
\bibitem{romaappr}
M.~Testa,
in Seoul 1997, Recent developments in nonperturbative 
quantum field theory p.~114, hep-lat/9707007 and references therein. 
\bibitem{fachin}
S.~Fachin and  C.~Parrinello,
Phys. Rev. {\bf D44} (1991) 2558.
\bibitem{parrisoft}
D.~S.~Henty, O.~Oliveira, C.~Parrinello and S.~Ryan, 
Phys. Rev. {\bf D54} (1996) 6923. 
\bibitem{giusti}
L. Giusti,
Nucl. Phys. {\bf B498} (1997) 331.
\bibitem{books}
C.~Itzykson and J.B.~Zuber,
Quantum Field Theory, Mc.Graw-Hill, New York (USA) 1980.
J.~Zinn-Justin,
Quantum Field Theory and Critical Phenomena, 
Oxford University Press, Oxford (UK) 1989.
\bibitem{herbert}
H.~ Neuberger,
Phys. Lett. {\bf B175} (1986) 69; Phys. Lett. {\bf B183} (1987) 337. 
\bibitem{pisa}
L.~Giusti, M.~L.~Paciello, S.~Petrarca, B.~Taglienti, 
presented at Lattice99, June 29 - July 3 1999, Pisa, Italy to appear 
in the proceedings, hep-lat/9910012. 
\bibitem{Gribov}
V. N. Gribov,
Nucl. Phys. {\bf B139} (1978) 1.
For a up to date discussion on the lattice see P.~van Baal,
Talk given at NATO Advanced Study Institute on Confinement, 
Duality and Nonperturbative Aspects of QCD, Cambridge, England, 
23 Jun - 4 Jul 1997. Cambridge 1997, Confinement, duality, and 
nonperturbative aspects of QCD 161-178, hep-th/9711070. 
\bibitem{noigauge}
L. Giusti, M.L. Paciello, S. Petrarca, B. Taglienti and M. Testa,
Phys. Lett. {\bf B432} (1998) 196. 
\bibitem{teorico}
A  perturbative study on the lattice  with periodic boundary conditions
is in preparation.
\end{thebibliography}
\end{document}